\documentclass[twocolumn,showpacs,preprintnumbers,amsmath,amssymb]{revtex4}

\usepackage{graphicx}
\usepackage{dcolumn}
\usepackage{bm}
\usepackage{hyperref}

\def\be{\begin{equation}}
\def\ee{\end{equation}}

\begin{document}

\preprint{hep-th/0608151}

\title{A Holographic Prediction of the Deconfinement Temperature}
 
\author{Christopher P. Herzog}
\affiliation{
Physics Department,
University of Washington \\
Seattle, WA  98195-1560
}%

\date{August 25, 2006}

\begin{abstract}
We argue that deconfinement in AdS/QCD models occurs via
a first order Hawking-Page type phase transition between
a low temperature thermal AdS space and a high temperature
black hole.
Such a result is consistent
with the expected temperature independence, to leading order in 
$1/N_c$, 
of the meson spectrum and spatial Wilson loops below the 
deconfinement temperature.  As a byproduct, we
obtain model dependent deconfinement temperatures $T_c$ in the hard and soft wall
models of AdS/QCD. Our result for $T_c$ in the soft wall model
is close to a recent lattice prediction.
\end{abstract}

\maketitle

\section{\label{sec:level1}Introduction}

Since the discovery of the Anti de Sitter space -- Conformal Field Theory (AdS/CFT) 
correspondence \cite{jthroat, GKP, EW} relating
type IIB string theory on $AdS_5 \times S^5$ and ${\mathcal N}=4$ super Yang-Mills (SYM) theory,
many have hoped that generalizations might yield deep insights into QCD and the nature of confinement.  Indeed, many qualitative advances in understanding have occurred.  The paper
\cite{thermalEW} studied confinement in ${\mathcal N}=4$ SYM by placing the gauge theory on 
a sphere and related deconfinement to a Hawking-Page phase
transition in the dual gravitational description.  The papers \cite{KlebanovStrassler, MaldacenaNunez, PolchinskiStrassler} realized confinement in related supersymmetric field theories by finding gravitational descriptions that capped off the geometry in a smooth way in the infrared at small radius.  
The gravitational descriptions in these last three papers are cumbersome, but the geometric insight is clear: cutting off the geometry at small radius produces confinement in the dual gauge theory.  
Based on this insight, Ref.~\cite{PSscatter} studied a far simpler model: $AdS_5$ where the small radius region is removed.  While such a removal is brutal, subsequent work has shown that one gets realistic, semi-quantitative descriptions of low energy QCD \cite{Erlich, daRoldPomarol}.

In this letter, we perform a semi-quantitative analysis of the deconfinement phase 
transition in these AdS/QCD models.  In particular, we consider both the hard wall model of 
Refs.~\cite{Erlich, daRoldPomarol} and also the soft wall model of Ref.~\cite{Karch}
where 
the authors study
a more gentle infrared truncation of $AdS_5$ induced by a dilaton-like field.  This soft wall model has the advantage of producing mesons with a stringy mass spectrum (as compared with the free particle
in a box spectrum of the hard wall model).

We find the deconfinement phase transition corresponds
to a Hawking-Page \cite{HawkingPage} 
type first order transition between thermal AdS at low temperature 
and an asymptotically AdS geometry
containing a black hole at high temperature.
A careful look at the gravitational free energies of the cut-off thermal AdS and the black hole solution reveal that the cut-off thermal AdS is stable for a range of temperatures where the black hole horizon radius would appear inside the AdS cavity.

Perhaps based on the observation that $T_c \to 0$ for ${\mathcal N}=4$
SYM on a sphere 
as the radius of the sphere goes to infinity, 
many authors have assumed (see for example \cite{Ghoroku, BoschiFilho, Andreev}) that the black hole
phase in 
these hard and soft wall models is always stable.  
This assumption 
leads to physics inconsistent with our expectations
of large $N_c$ gauge theories, as we discuss at the end of the letter.
While our argument for a phase transition makes certain subtle assumptions about
the gravitational action, the existence of this phase transition
is fully compatible with our field theory understanding.

At a time where increasingly researchers are trying to apply AdS/CFT inspired
models and calculations to experiment, investigating the consistency and universal
features of these models is of critical importance.  To cite two better known
examples, Ref.~\cite{viscosity} has attempted to use the low value of the viscosity
to entropy density ratio in these and related models to explain high values of the elliptic flow
in heavy ion collisions at RHIC.  The Refs.~\cite{heavyquark} attempt to measure
the energy loss rate of heavy quarks from AdS/CFT
to gain a better understanding of charm and bottom
physics at RHIC.  In the absence of a gravity dual for QCD, one
approach to experiment is to seek out universal behavior in the gravity duals
we do understand;
in both models we study, we see evidence for a first
order phase transition.  While AdS/CFT remains a conjecture, it is important 
to check that these dual models are consistent with field theory understanding;
the phase transition we find is fully
compatible with our large $N_c$ field theory expectations, as we argue at the end.

As a byproduct of our analysis, we relate the mass of vector mesons to the deconfinement
temperature.  The vector mesons correspond to cavity modes in the cut-off AdS.
By matching the mass of the lightest vector meson to experimental data, we can fix the infrared
cut-off scale.  The Hawking-Page analysis then relates this cut-off scale to the deconfinement temperature.   Our prediction of $T_c \approx 191 \mbox{ MeV}$ for the soft wall model 
is amusingly close to one recent lattice 
prediction \footnote{$T_c = 191 \mbox{ MeV}$ comes from \cite{lattice}.
Note however a more recent result \cite{Wuppertal} 
arguing that the crossover transition
is broad of order 150 to 170 MeV
and depends on the observable chosen as order parameter.
}.

We begin in Section II by performing an analysis of the Hawking-Page phase transition for the hard and soft wall AdS/QCD models.  In Section III, we review some results of \cite{Erlich, daRoldPomarol, Karch} for vector meson masses in these models to extract a prediction for the deconfinement temperature.  Section IV concludes with remarks about temperature independence of equilibrium 
quantities in the confining phase at a large number of colors $N_c$.

\section{Hawking-Page Analysis}

\subsection{The Hard Wall Model}

We establish a relationship between the confinement
temperature and the infrared cut-off for the hard wall model assuming
that the thermodynamics is governed by the gravitational part of the
action.  The assumption is justified at large $N_c$ where
the gravitational part scales as $N_c^2$ while the contribution
from the mesons we consider later scales only as $N_c$.

We consider a gravitational action of the form
\be
I = - \frac{1}{2\kappa^2} \int d^5x \sqrt{g}  \left(R + \frac{12}{L^2} \right) \ .
\label{Ihard}
\ee
The gravitational coupling scales as $\kappa \sim g_s \sim 1/N_c$.
There are two relevant solutions to the equations of motion.  The first is
cut-off thermal AdS with a line element
\be
ds^2 = L^2 \left( \frac{ dt^2 +  d\vec x^2 + dz^2}{z^2} \right) 
\ee
where the radial coordinate extends from the boundary of AdS $z=0$ to a cut-off
$z=z_0$.
The second solution is cut-off AdS with a black hole with the line
element:
\be
ds^2 = \frac{L^2}{z^2} \left( f(z) dt^2 + d \vec x^2 + \frac{dz^2}{f(z)} \right) \ ,
\label{blackhole}
\ee
where $f(z) = 1- (z/z_h)^4$.  The Hawking temperature of the black hole solution
is $T = 1/ (\pi z_h)$.  

In both cases, we continued to Euclidean signature with a compact time direction.  In the black hole case, the periodicity is enforced by regularity of the metric near the horizon, $0 \leq t < \pi z_h$.  In thermal AdS, the periodicity of $t$ is not constrained.

In either case, the curvature of the solution is $R= -20 / L^2$ and so on-shell,
the gravitational action becomes
\be
I = \frac{ 4}{L^2 \kappa^2} \int d^5 x \sqrt{g} \ ,
\label{onshell}
\ee
i.e.~the volume of space-time times a constant \footnote{
There is a surface term, but, as noted by
\cite{HawkingPage}, the surface term contribution from $z \to 0$ 
vanishes for these space-times.  In principle, there could be a surface term
from $z=z_0$ as well.  In the spirit of \cite{KlebanovStrassler, MaldacenaNunez}
where space-time ends at $z=z_0$ without a boundary, it seems best
to set the Gibbons-Hawking term at $z=z_0$ to zero by hand.}.

The value of $I$ for both space-times is infinite, so we regularize by integrating 
up to an ultraviolet cut-off $z=\epsilon$.  
(We divide out by the trivial infinity related to the integral over the
spatial components $\vec x$ of the metric.)
For thermal AdS, the regularized action density becomes
\be
V_1(\epsilon) = \frac{4 L^3}{\kappa^2} \int_0^{\beta'} dt \int_{\epsilon}^{z_0} dz \, z^{-5} \ ,
\ee 
while for the black hole in AdS, the density is
\be
V_2(\epsilon) = \frac{4 L^3}{\kappa^2} \int_0^{\pi z_h} dt \int_{\epsilon}^{\mbox{\footnotesize min}(z_0, z_h)} dz \, z^{-5} \ .
\ee
These $V_i$ are free energy densities in the field theory.

We compare the two
geometries at a radius $z=\epsilon$ where the periodicity in the time direction is locally
the same.  In other words, $\beta'  = \pi z_h \sqrt{f(\epsilon)}$.  After this adjustment,
\begin{eqnarray}
\Delta V &=& \lim_{\epsilon \to 0} \left(V_2(\epsilon) - V_1(\epsilon) \right) \nonumber \\
&=& 
\left \{
\begin{array}{lc}
\frac{L^3 \pi z_h}{\kappa^2} 
\frac{1}{2 z_h^4}  & z_0 < z_h \\
\frac{L^3 \pi z_h}{\kappa^2} 
\left(\frac{1}{z_0^4} - \frac{1}{2 z_h^4}  \right) & z_0 > z_h \\
\end{array}
\right. \ .
\end{eqnarray}
When $\Delta V$ is positive (negative), thermal AdS (the black hole) is stable.  
Thus the Hawking-Page phase transition occurs at a temperature corresponding
to $z_0^4 = 2 z_h^4$, or 
\be
T_c = 2^{1/4} / (\pi z_0) \ .
\label{hardT}
\ee
As the temperature increases, thermal AdS becomes unstable and the black hole becomes stable.  
At $T_c$, the black hole horizon forms inside the AdS cavity, between the boundary
and the infrared cut-off, at a radius $z_h<z_0$.  

\subsection{The Soft Wall Model}

The calculation for the soft wall model of \cite{Karch} is similar to the hard wall model calculation
described previously.  Certain features 
require additional
explanations and assumptions.  In place of (\ref{Ihard}), we have the action
\be
I = -\frac{1}{2\kappa^2} \int d^5 x \sqrt{g} e^{-\Phi}  \left(R + \frac{12}{L^2} \right)
\ee
where $\Phi = c z^2$ is a dilaton like field taken to have nontrivial expectation value.
This dilaton field is assumed not to affect the gravitational
dynamics of our theory.  As in \cite{Karch}, we assume that AdS space solves the
equations of motion for the full theory.  We make the additional assumption that
the black hole in AdS (\ref{blackhole}) also satisfies the equations of motion.  
The on-shell action is then (\ref{onshell}) scaled by a dilaton dependent factor:
\be
I = \frac{ 4}{L^2 \kappa^2} \int d^5 x \sqrt{g} e^{-\Phi} \ .
\label{onshell2}
\ee
To trust this set of assumptions, we should construct an explicit supergravity
background with these properties, something we have not done.  
Such a solution may exist.  In string frame, the dilaton kinetic factor has the wrong sign,
and it is conceivable one may construct a nontrivial solution with a trivial stress energy tensor. 
For example the dilaton-tachyon system considered by \cite{RazamatBergman} has
precisely such a solution with a quadratic dilaton and linear tachyon but also breaks Lorentz invariance
\cite{Karchpersonal}.
Regardless of its quantitative significance, qualitatively our answer
must be right because it conforms with our large $N_c$ field theory expectations, as we
argue at the end.

From this on-shell action (\ref{onshell2}), we calculate the regularized action densities for thermal AdS:
\be
V_1(\epsilon) = \frac{4 L^3}{\kappa^2} \int_0^{\beta'} dt \int_{\epsilon}^{\infty} dz \, z^{-5} e^{-c z^2} \ ,
\ee
and for the black hole solution
\be
V_2(\epsilon) = \frac{4 L^3}{\kappa^2} \int_0^{\pi z_h} dt \int_{\epsilon}^{z_h} dz \, z^{-5} e^{-c z^2}\ .
\ee
Choosing $\beta'$ as in the hard wall model, $\beta' = \pi z_h \sqrt{f(\epsilon)}$,
the end result is 
\begin{eqnarray}
\Delta V &=& \lim_{\epsilon \to 0} (V_2 - V_1)  \\
&=& \frac{\pi L^3}{\kappa^2 z_h^3} \Big(
e^{-c z_h^2} \left( -1  + c z_h^2  \right) +\frac{1}{2}
+  c^2 z_h^4 \mbox{Ei}( {-c} z_h^2)
\Big) \ .  \nonumber 
\end{eqnarray}
Here, $\mbox{Ei}(x) \equiv - \int_{-x}^\infty e^{-t}/t \, dt$.  Numerically,  
there will be a phase transition from
thermal AdS to the black hole solution when $c z_h^2 = 0.419035\ldots$,
or
\be
T_c = 0.491728 \sqrt{c} \ .
\label{softT}
\ee
For small temperatures (large $z_h$), $\Delta V \to L^3 \pi / (2\kappa^2 z_h^3) > 0$ and 
thermal AdS is stable.  For large temperatures (small $z_h$), $\Delta V \to -L^3 \pi / (2\kappa^2 z_h^3) < 0$
and the black hole solution is stable.

\begin{figure}
\includegraphics[width=3in]{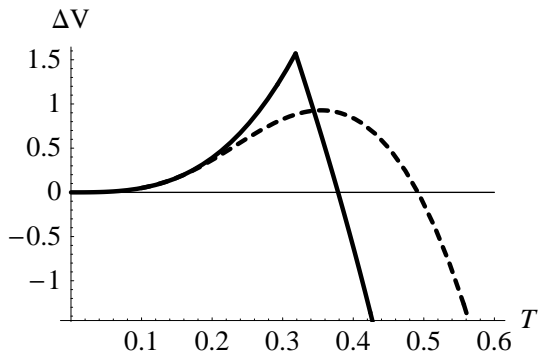}
\caption{\label{myfig} 
The solid line is the free energy difference in the hard wall model, the dashed line the difference
in the soft wall model.  The graph was made in units where $L^3/\kappa^2=c=z_0=1$.
}
\end{figure}

\section{Vector Mesons and Matching QCD}

In the previous section, we found expressions (\ref{hardT}) and (\ref{softT}) 
that related the deconfinement
temperature to, in one case, the infrared hard wall cut-off $z_0$ and, in the other, 
the soft wall parameter
$c$.  Here we review results from Refs.~\cite{Erlich, daRoldPomarol, Karch, Andreevtwo} 
that relate $z_0$
and $c$ to the spectrum of vector mesons in QCD and thus get a relation between the mass of the
lightest vector meson and the deconfinement temperature.


In the hard and soft 
wall cases, Refs.~\cite{Erlich, daRoldPomarol, Karch} 
model vector mesons as cavity modes
of a vector field in this modified AdS space.  Choosing a radial gauge where $V_z=0$, 
these vector fields 
$V_\mu(x,z) = V_\mu(q,z) e^{i q \cdot x}$ satisfy the equation of motion
\be
\partial_z \left( \frac{1}{z} e^{-\Phi} \partial_z V_\mu(q,z) \right) - \frac{e^{-\Phi} q^2}{z} V_\mu(q,z) = 0 \ ,
\label{vectormode}
\ee
where $\Phi=0$ in the hard wall model and $\Phi=cz^2$ in the soft wall model.

In the hard wall case, normalizable boundary conditions
at $z=0$ determine that the solutions are Bessel functions:
$
V_\mu(q,z) \sim z J_1 (m z)
$, where $m^2 = -q^2$.
Applying Neumann boundary conditions at the cut-off $z=z_0$, 
one finds only a discrete set of eigenmodes corresponding
to discrete choices of $q$ which satisfy $J_0(m_i z_0) = 0$.  The first zero of $J_0(x)$ occurs
at $x = 2.405\ldots$, implying the lightest $\rho$ meson has a mass $m_1=2.405/z_0$.  
Experimentally, the lightest $\rho$ meson has a mass of $m_1 = 776 \mbox{ MeV}$. 
Thus, we conclude that  
$z_0 = 1/(323 \mbox{ MeV})$.

We now can make a prediction for the deconfinement temperature in this hard wall model:
\be
T_c = 2^{1/4} / (\pi z_0) \approx 0.1574 \, m_\rho = 122  \mbox{ MeV} \ ,
\ee
a low number compared to new lattice estimates \cite{lattice, Wuppertal}.


In the soft wall model, 
the relevant solution to this differential equation (\ref{vectormode}) involves Laguerre polynomials:
$
V_\mu(q,z) \sim z^2 L_n^1 (c z^2)
$
where the allowed values of $q$ are $-q^2 = 4 n c$ ($n \in \mathbb{Z}^+$). 
 Matching to the lightest
$\rho$-meson, we find $\sqrt{c} = 338 \mbox{ MeV}$.  
Our prediction for the deconfinement
temperature in the soft wall model is thus
\be
T_c = 0.4917 \sqrt{c} \approx 0.2459 \, m_\rho = 191 \mbox{ MeV}
\ee
which is a current lattice prediction \cite{lattice}.  Because 
phase transitions are sensitive to the density of states, perhaps the fact that the soft wall model
produces a more realistic meson spectrum is related to this improved prediction compared with the hard wall model.  

\section{Discussion}

We argue that the 
stability of thermal AdS at low temperatures and the presence of a first
order phase transition in these soft and hard wall models of 
QCD 
is consistent with large $N_c$ field theory expectations.
Recall at 
large $N_c$, the confining low temperature phase has $O(1)$ entropy density, discrete meson and glueball spectra, and vanishing expectation 
value for the Polyakov loop (a time like Wilson loop).  The deconfined, high temperature phase
has $O(N_c^2)$ entropy density, temperature dependent spectral
densities, and a non-zero expectation value for the Polyakov loop.

From these properties, it follows that in soft wall models,
the black hole configuration cannot be stable for all temperatures. 
The soft wall model is intended to be a model of QCD which experiences
a phase transition at $T_c>0$, but the presence of a horizon
at $T=0$ indicates that $T_c=0$. 
In the gravity dual, the presence of a horizon introduces $O(N_c^2)$ degrees of freedom.
Mesons \cite{Karch}
and glueballs in this soft-wall model correspond to discrete cavity modes, 
but the presence of a horizon leads to loss of probability density into the 
black hole and smears out the spectrum.  Moreover, with a horizon, there is
no obvious topological reason for the Polyakov loop to vanish \cite{thermalEW}.  
In the hard wall model,
the sharp cut-off at $z=z_0$ can completely hide the horizon and the preceding horizon
dependent arguments fail.  However, even if the cut-off at $z_0<z_h$ completely cloaks the
horizon, 
the metric far from the horizon is still different
in the black hole background and leads to 
spatial Wilson loops and a mass spectrum for mesons and glueballs
inconsistent with our large $N_c$
expectations, as we now argue.

 In the confined phase, spatial Wilson loops 
have an expectation value that to leading order in $1/N_c$ is 
temperature independent.  From the field theory perspective, a spatial Wilson loop produces
a sheet of flux that sits
at a point in the compactified time direction.  Temperature dependence
can only come from fluctuations that are large compared to the inverse temperature
scale, wrap around the compactified time direction, and lead to self intersections of the flux sheet.
These self intersections are suppressed by a power of $1/N_c$.
From the gravitational perspective, these spatial Wilson loops correspond to Euclidean
string world sheets that droop from the boundary of AdS toward the center.  The area law comes from
the fact that for a large enough boundary, most of the string world sheet lies along the cut-off.
If the black hole configuration were always stable, even though the horizon is hidden behind the cut-off, the string would experience temperature
dependent curvature corrections that alter its effective tension.  Instead, since thermal AdS is thermodynamically preferred, the expectation value will be temperature independent.

Next, consider the mesons
and glueballs which in
these AdS/QCD models correspond to cavity modes
 \cite{Erlich, daRoldPomarol, Karch, glueball}.  Again, if the black hole solution
were always stable, even if the event horizon were effectively hidden by the cut-off, temperature
dependent curvature corrections would appear in the mass spectrum.  Such corrections
are not expected from the point of view of large $N_c$ field theory.
In the confining phase, the interaction cross sections of mesons and glue balls 
are $1/N_c$ suppressed.  
From chiral perturbation theory, for example, 
Ref.~\cite{Ioffe} demonstrated that temperature dependent
corrections to meson masses involve diagrams with at least two pions in the intermediate
channel.  Since the decay width to pions is already $1/N_c$ suppressed, 
the mass corrections must be as well.   
More formally, we could integrate out the fermions from our theory and re-express
the mass corrections for mesons in terms of sums of non-local operators, for example, 
the spatial Wilson loops discussed above which have
$1/N_c$ suppressed temperature corrections \cite{orbifold}
\footnote{
I would like to thank Larry Yaffe for discussion of these issues.
}.

In conclusion, we emphasize that the formation of a black hole in these
AdS/QCD models does not happen at $T=0$, but is instead dual to a first order
deconfinement phase transition at a finite temperature $T_c>0$. A black hole
at $T=0$ would lead to ${\mathcal O}(N_c^2)$ entropy density, a nonvanishing
Polyakov loop, leading order in $1/N_c$ temperature corrections to spatial
Wilson loops, and other effects inconsistent with our expectations for 
the confining phase of a large $N_c$ gauge theory.

\vskip 0.1in 

\begin{acknowledgments}
I would like to thank Andreas Karch, Berndt Mueller, Steve Sharpe, 
Aleksi Vuorinen, and Larry Yaffe for useful discussions and for encouraging me to write up these observations.
This work was supported in part by the U.S. Department of Energy under Grant No.~DE-FG02-96ER40956.
\end{acknowledgments}

\end{document}